\documentclass[conference]{IEEEtran}
\IEEEoverridecommandlockouts
\usepackage{cite}
\usepackage{amsmath,amssymb,amsfonts}
\usepackage{algorithmic}
\usepackage{multirow}
\usepackage{graphicx}
\usepackage{booktabs}
\usepackage{textcomp}
\usepackage{color}
\usepackage{xcolor}
\usepackage{comment}
\usepackage{enumitem}
\usepackage{hyperref}
\usepackage{caption}
\usepackage{subcaption}
\usepackage{esdiff}
\usepackage{todonotes}
\usepackage{soul}
\usepackage[mathscr]{eucal}
\def\BibTeX{{\rm B\kern-.05em{\sc i\kern-.025em b}\kern-.08em
    T\kern-.1667em\lower.7ex\hbox{E}\kern-.125emX}}

\DeclareMathOperator*{\argmin}{arg\,min}
\newcommand{\etal}{\textit{et al}.~}
    
\usepackage[normalem]{ulem}
\definecolor{darkgreen}{cmyk}{0.90,0,0.80,0.20}
\definecolor{darkred}{cmyk}{0, 1, 1, 0.10}
\definecolor{lightblue}{cmyk}{1, 0.70, 0, 0}

\usepackage{color}
\usepackage{soul}

\begin{document}

\title{A Closer Look at Evaluating the Bit-Flip Attack Against Deep Neural Networks \\{\LARGE\textit{(extended version)}}}

\author{
\IEEEauthorblockN{Kevin Hector\IEEEauthorrefmark{1}\IEEEauthorrefmark{2}, Pierre-Alain Moëllic\IEEEauthorrefmark{1}\IEEEauthorrefmark{2}, Mathieu Dumont\IEEEauthorrefmark{1}\IEEEauthorrefmark{2}, Jean-Max Dutertre\IEEEauthorrefmark{3}}
\IEEEauthorblockA{\IEEEauthorrefmark{1}CEA Tech, Centre CMP, Equipe Commune CEA Tech - Mines Saint-Etienne, F-13541 Gardanne, France}
\IEEEauthorblockA{\IEEEauthorrefmark{2}Univ. Grenoble Alpes, CEA, Leti, F-38000 Grenoble, France \\ 
\emph{\{kevin.hector,pierre-alain.moellic, mathieu.dumont\}@cea.fr}}
\IEEEauthorblockA{\IEEEauthorrefmark{3}Mines Saint-Etienne, CEA, Leti, Centre CMP, F-13541 Gardanne, France \\
\emph{dutertre@emse.fr}}
}

\maketitle

\begin{abstract}
Deep neural network models are massively deployed on a wide variety of hardware platforms. This results in the appearance of new attack vectors that significantly extend the standard attack surface, extensively studied by the adversarial machine learning community. One of the first attack that aims at drastically dropping the performance of a model, by targeting its parameters (weights) stored in memory, is the \textit{Bit-Flip Attack} (BFA). In this work, we point out several evaluation challenges related to the BFA. First of all, the lack of an \textit{adversary's budget} in the standard threat model is problematic, especially when dealing with physical attacks. Moreover, since the BFA presents critical variability, we discuss the influence of some training parameters and the importance of the model architecture. This work is the first to present the impact of the BFA against fully-connected architectures that present different behaviors compared to convolutional neural networks. These results highlight the importance of defining robust and sound evaluation methodologies to properly evaluate the dangers of parameter-based attacks as well as measure the real level of robustness offered by a defense.
\end{abstract}


\section{Introduction}
An important trend in modern artificial intelligence is the deployment of deep neural networks in a wide variety of hardware platforms (e.g., FPGA, 32bit microcontrollers). This large-scale deployment raises major security concerns about their integrity, confidentiality, privacy and availability. Among them, a significant amount of works deals with adversarial examples~\cite{szegedy2013intriguing} that are perturbations of the inputs at inference time that fool the predictions of a model.  

Since the attack surface also encompasses the implementation of models, recent works show that attacking model parameters (parameter-based attacks) or inference instructions (e.g., bias values~\cite{liu2017fault} or activation functions~\cite{breier2018practical} with fault injection) are worrying attack vectors against the model integrity. 

A milestone attack proposed by Rakin \etal\cite{Rakin_2019_ICCV}, called Bit-Flip Attack (hereafter BFA), targets the parameters (also called \textit{weights}) of a neural network stored in memory.
The BFA (or variants \cite{rakin2021tbfa,park2021zebra}) aims at flipping some bits of weights in order to globally decrease the performance of the target model. The bits are picked according to their influence on the loss in a fashion close to gradient-based adversarial examples crafting method. 

Following~\cite{Rakin_2019_ICCV}, several works proposed additional experiments and analysis~\cite{rakin2021tbfa} and defenses~\cite{He_2020_CVPR, cherupally2021leveraging}.
Most of the works related to the BFA are based on simulations but refer to practical means to perform fault injection since there exists a rich state-of-the-art on this field, mainly in the context of cryptographic modules~\cite{5560194} and recently with a growing interest for embedded machine learning models~\cite{breier2018practical, 9595075}. In~\cite{yao2020deephammer}, the BFA has been practically demonstrated using a single-sided RowHammer method against models stored in DRAM. 

Even if parameter-based attacks are still in its infancy, a parallel has to be drawn with input-based attacks. An impressive number of adversarial example attacks and defenses have been proposed and a significant part raises very critical evaluation issues (pointed out in reference works~\cite{athalye2018obfuscated,carlini2019evaluating,carlini2017adversarial,carlini2017towards,tramer2020adaptive,zimmermannincreasing}) that alter the confidence on the level of robustness a model can claimed. Being an essential concern, and regarding the rapid evolution of modern deep neural network models, parameters-based attacks also need further analysis and sound evaluation methodologies. In this context, our contributions are as follow:
\begin{itemize}
    \item We question the relevance of the criteria used to measure the BFA since  \cite{Rakin_2019_ICCV}, that consists in reaching a random-guess level, because it misrepresents the evaluation of the attack, especially in the context of fault injection.
    \item Previous experiments show problematic high variance. We observe that the BFA can be dependent on training parameters, that should be taken into account for evaluation, and highlight the importance of the model architecture with the first results for fully-connected networks.
    \item Experimentally, for models that do not have the same properties than typical convolutional models, we show that the standard BFA can be significantly non-optimal compared to a very simple variation and therefore can lead to a false sense of robustness.   
\end{itemize}

For reproducibility purpose, setups, codes of our experiments as well as detailed results are publicly available\footnote{\url{https://gitlab.emse.fr/securityml/closerlook-bfa}}.

\section{Preliminaries and notations}
\subsection{Formalism}

In this work, we consider a neural network model that performs a classification task where input-label pairs $(x,y) \in \mathcal{X} \times \mathcal{Y}$ are sampled from a distribution $\mathcal{D}$. $\left|\mathcal{Y}\right| = C$ is the cardinality of the labels space. 
The neural network model $M_W : \mathcal{X} \rightarrow \mathcal{Y}$, with parameters $W$ (also referred as \textit{weights}), classifies an input  $x \in \mathcal{X}$ to a label $M_W(x) \in \mathcal{Y}$. For the sake of readability, the model $M_W$ is simply noted as $M$, except when necessary. $M$ is composed of several layers that are successively connected. A layer performs a weighted sum of its inputs (i.e., the previous layer or the inputs data for the first one), then, a non-linear mapping thanks to an activation function $\phi$.  
$\mathcal{L}\big(M_W(x),y\big)$ is the loss function (typically the cross entropy for classification task) for $M_W$ that quantifies the error between the prediction $\hat{y} = M_W(x)$ and the groundtruth label $y$. The training process aims at finding the best parameters that minimize the loss on the training dataset (known as the \textit{Empirical Risk Minimization}, Equation~\ref{ml_definition}).

\begin{equation}
W^* = \argmin_W\big(
\mathop{\mathbb{E}}_{x,y\sim \mathcal{D}}\big[\mathcal{L}(M_{W}(x);y)\big] \big)
\label{ml_definition}
\end{equation}

The optimization process to find the best parameters is based on the gradient of the loss w.r.t. each parameter $w$~\cite{goodfellow2016deep}. A parameter $w$ is updated in an iterative way at each \textit{mini-batch} (a small fragment of the training set) with Equation~\ref{eq_parameter_update}:

\begin{equation}
w^{t+1} = w^{t} - \lambda \nabla_{w}\mathcal{L}(M(x),y))
\label{eq_parameter_update}
\end{equation}

$\lambda$ is the \textit{learning rate}. The weights are initialized with~--~typically~--~a normal or uniform distribution. The \textit{learning rate} and the \textit{weight initialization} are studied in section~\ref{training_param}. 

\subsection{Models and datasets}
\label{experimental_setup}

Following most of the works addressing parameter-based attacks, we use CIFAR-10 that is composed of 60,000 color images (32x32) divided in 10 classes, including 10,000 samples for testing.
We also use MNIST, composed of 60,000 training and 10,000 testing black and white digits (28x28), because different types of neural network architectures are easily trainable on this dataset.  

As in~\cite{Rakin_2019_ICCV} and~\cite{He_2020_CVPR}, we apply the BFA on two popular convolutional neural network architectures (hereafter, \textbf{CNN}): \textbf{ResNet-20}~\cite{he2016deep} and \textbf{VGG-11}~\cite{Simonyan15}.
 We add two custom models: (1) a fully-connected model (multi-layer perceptron, hereafter \textbf{MLP}) composed of four fully-connected layers with 512, 256, 128, 10 neurons and ReLU activation functions; (2) a variant of MLP with an additional convolutional layer as the first layer (32 filters of size 3x3) that we refer as \textbf{C-CNN}.
 
Similarly to the previous works we mentioned, our models are trained with 8 bits quantization aware-training since crushing a full-precision model is as easy as attacking the most significant bit of the exponent part of a single weight~\cite{Rakin_2019_ICCV} (value explosion). The accuracy of the models as well as the detailed training parameters for each experiment is proposed in appendix (\ref{annex_training_setup}, \ref{annex_model_accuracy}).

\section{Bit-Flips Attack} \label{BFA_section}
\begin{figure}[t]
\centerline{\includegraphics[width=0.49\textwidth]{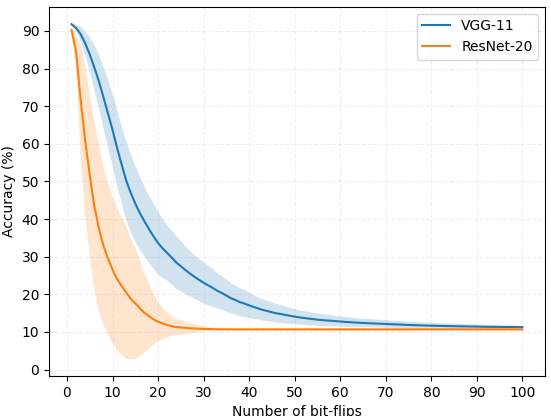}}
\caption{BFA results on ResNet-20 and VGG-11. Mean (stdev) on five models and attack sets.}
\label{fig_basic_models}
\end{figure}

\subsection{Original threat model}

The BFA~\cite{Rakin_2019_ICCV}~identifies and flips the most sensitive bits of the parameters of a model $M_W$ in order to drastically decrease its accuracy. From~\cite{Rakin_2019_ICCV}, the associated threat model is as follows:
\begin{itemize}
\item\textit{Adversary's knowledge}: the BFA is a white-box attack, the attacker needs a perfect knowledge of $M_{W}$ to compute gradients of the loss according to the weights $\nabla_w\mathcal{L}$.
\item\textit{Adversary's goal}: as presented in~\cite{Rakin_2019_ICCV} and widely reused in other works, the goal is to decrease the accuracy of $M$ until an almost random-guess level ($\approx1/C$). Rakin~\etal~\cite{Rakin_2019_ICCV} used an accuracy of 0.11 for CIFAR-10.
\item\textit{Adversary's budget}: Interestingly, the maximum number of bit-flips allowed is hardly ever mentioned, i.e. the adversary is able to perform as many faults as needed to reach the random-guess objective. 
\end{itemize}

We discuss this threat model in section~\ref{discussion_threat_model}. 
\subsection{Attack principle}

The BFA starts with a Progressive Bit Search method (PBS)~\cite{Rakin_2019_ICCV}~that identifies the most sensitive bits, followed by the BFA attack that flips the bits previously identified. The two methods are performed iteratively until reaching the adversary's goal. That attack is performed on a small set from the train set, called the \textit{attack set}. 

The PBS alternates, for each iteration, an in-layer and cross-layer search. First, the in-layer search selects the best bit in the layer $l$ by ranking the gradients of the bit $b$ w.r.t. the loss: $\nabla_b\mathcal{L}$. After the most sensitive bit of each layer is found, each bit is flipped (and then restored) to measure the performance loss after this (and only this) bit-flip. After processing all the layers, the one with the maximum loss is selected and the corresponding bit-flip is~--~this time~--~permanently performed. The bit-flip is realized along the gradient ascendant w.r.t. the loss \( \mathcal{L} \) as defined in~\cite{Rakin_2019_ICCV} with Eq.~\ref{eq1}: 

\begin{equation}
\label{eq1}
\begin{split}
    &m= b\oplus \big(sign(\nabla_b\mathcal{L})/2 + 0.5\big)\\
    &\hat{b} = b \oplus m
\end{split}
\end{equation}

With $\hat{b}$, the bit after bit-flip. Interestingly, BFA follows the principle of most adversarial example crafting methods by relying on the direction that may increase (for untargeted attack) or decrease (for targeted attack) the loss. Thus, Equation~\ref{eq1} can be seen as a variant of the Fast Gradient Sign Method (FGSM) \cite{goodfellow2014explaining}. However, this gradient heuristic is not said to be the most efficient and, very recently, authors from~\cite{lee2022sparsebfa} show that a Taylor’s expansion-based heuristic ($|w.\nabla_{w}\mathcal{L}|$) is more efficient than the gradient ($|\nabla_{w}\mathcal{L}|$) for sparse networks (i.e., models compressed by a high pruning rate).

\subsection{Standard BFA performance (ResNet-20, VGG-11)}
\label{standard_bfa_perf}
For all the experiments of this paper, the BFA configuration is as follows: the attack dataset is a random sampling of 256 images from the train set
and each architecture is trained five times (different seeds for the weight initialization) and each trained model is attacked five times (i.e., we perform a total of 25 BFA for each architecture). The accuracy of the five models for each architecture is provided in Table~\ref{tab_accuracy} in the Appendix. 

Fig.~\ref{fig_basic_models} shows the incremental drop of performance (accuracy) with respect to the number of bit-flips for ResNet-20 and VGG-11. 
 These results are averaged over five training and five attack trials (area around the curve is the standard deviation). For these experiments, we used the same training parameters and models as in~\cite{He_2020_CVPR} (and publicly available, see Section~\ref{comment_sota}).
 
 The shape of the curves indicates a high efficiency of the first bit-flips that strongly alter the accuracy, then it needs an accumulation of many errors to reach the random-guess level. We analyse this phenomenon in the next section.
 
 \begin{table}[th]
\caption{Distribution over the layers and contribution of the bit-flips for ResNet-20. Each \textit{stage} gathers 6 convolutional layers.}
\centering
\begin{tabular}{ccc}
\toprule
Layer / Stage & Dist. bit-flips (\%)& Damage (\%) \\

\midrule
Conv. 1 & 21.35 & 23.87 \\
Stage 1 & \textbf{28.34} & \textbf{49.1} \\
Stage 2 & 23.96 & 20.53 \\
Stage 3 & 21.55 & 5.55 \\
Dense 1 & 4.80 & 0.95 \\
\bottomrule
\end{tabular}
\label{tab_distrib_resnet20_sumup}
\end{table}

Table~\ref{tab_distrib_resnet20_sumup} summarizes the distribution of the bit-flips for ResNet-20 ( Table~\ref{tab_distrib_resnet20} and \ref{tab_distrib_vgg11} in the Appendix provide all the detailed distribution for ResNet-20 and VGG-11).
For both models, we observe the same phenomenon with a concentration of the bit-flips in the first convolutional layer (for VGG-11, 22.27\% that represents 50.71\% of damage).
A first observation is that the closer the layer is to the model input, the higher the number of bit-flips is. Second, the bit-flips do not have the same contribution in the drop of accuracy: the more the bit-flip is performed in a layer close to the model input, the higher its contribution is. These results are coherent with \cite{He_2020_CVPR} and show that a bit-flip performed among the first layers has a strong impact because the perturbation is accumulated throughout the model.
\subsection{Sensitivity of the BFA}
\label{comment_sota}

The first line of Table~\ref{BFA_adversary_goal} is the result of the standard BFA for five VGG-11 models (simply named ``1", ...,``5") trained with different initial training seed (Normal distribution). The results are averaged over five attacks (i.e. five attack sets). The sensitivity w.r.t. the training initialization and the attack set is significant (e.g., a standard deviation of almost 52 for the model ``5") and show that, for this kind of models and the standard threat model, the BFA cannot be properly evaluated with a single model and attack set.  

The sensitivity of VGG-11 is critical in~\cite{He_2020_CVPR} with two experiments conducted on the same VGG-11 model in~\cite{He_2020_CVPR} that led to two different average number of bit-flips needed to reach the adversary's goal: 56.6 ± 35.2 and 16.4 ± 1.14 respectively\footnote{Fig.2 and Table 4 in~\cite{He_2020_CVPR}}. Our tests using the original source code confirm an average performance closer to 56 bit-flips on average rather than 16. This issue is mentioned in the public repository\footnote{\url{https://github.com/elliothe/BFA/issues/7}} of~\cite{He_2020_CVPR}. For reproducibility, the authors provide a model for which only 10 bit-flips are actually needed. However, we highlight the fact that this model is a special case with a very particular distribution of the weights: all the weights are close to zero except one weight per layer with an absolute value close to 3. With the effect of quantization, this highly unbalanced distribution leads to a condition that favors the attack by boosting the range of the perturbation.
We do not observe such unbalanced distribution for our five VGG-11 models as illustrated in Fig.~\ref{fig_weigths_distribution_8bits} (Appendix \ref{appendix_wd}).

In~\cite{Rakin_2019_ICCV} and \cite{He_2020_CVPR}, authors only considered CNN models (VGG-11 and ResNet-20). In our work, we also analyze a pure  fully-connected network (MLP), following recent modern architectures (e.g., Transformers~\cite{dosovitskiy2020image}, MLPMixer~\cite{tolstikhin2021mlp}) that reach state-of-the-art results and are the main topic of an important body of works in the machine learning community. Importantly, our results highlight a different behavior of this type of models, which we detail in the sections~\ref{training_param} and~\ref{focus_mlp}.  

\section{Adversary's goal and budget}
\label{discussion_threat_model}
\begin{table*}[t!]
\caption{Mean (standard deviation) of bit-flips over 5 attacks to reach 11\%, 25\%, 50\% and 75\% of accuracy, for 5 training runs (training seeds are in the Appendix).
}
\begin{center}
\resizebox{\linewidth}{!}{%
\begin{tabular}{ccccccccccc}

\toprule

\multirow{2}*{Acc goal (\%)}&\multicolumn{5}{c}{ResNet-20}&\multicolumn{5}{c}{VGG-11} \\
%\cline{2-5} 
 & \textbf{1}& \textbf{2} & \textbf{3} & \textbf{4} & \textbf{5} & \textbf{1} & \textbf{2} & \textbf{3} & \textbf{4} & \textbf{5}\\

\midrule

11~\cite{Rakin_2019_ICCV} & 20.6 (\textbf{5.08}) & 18.8 (\textbf{8.28}) & 21.4 (\textbf{4.49}) & 27.4 (\textbf{11.22}) & 7.0 (\textbf{2.09}) & 72 (\textbf{36.01}) & 55.4 (\textbf{20.92}) &  81.6 (\textbf{25.5}) & 42.6 (\textbf{9.02}) & 113.2 (\textbf{51.92}) \\
25 & 8.8 (1.83) & 8 (0.63) & 9.6 (1.02) & 12.4 (1.02) & 3.6 (0.49) & 14.2 (4.21) & 13.8 (2.71) &  16.0 (2.19) & 13.2 (1.47) & 19.6 (3.77) \\
50 & 6.2 (1.72) & 4.4 (0.49) & 5.4 (0.8) & 6.6 (0.8) & 2.4 (0.49) & 6.8 (0.75) & 7 (1.1) &  8.4 (1.36) & 6.6 (0.8) & 9(1.1) \\
75 & 3.4 (0.8) & 2.2 (0.4) & 3 (0.63) & 3.6 (0.49) & 1.6 (0.49) & 3 (0) & 3.2 (0.4) &  3.4 (0.49) & 3.2 (0.4) & 4 (0) \\

\bottomrule

\end{tabular}
}
\label{BFA_adversary_goal}
\end{center}
\end{table*}

For these experiments, we derive the setup from~\cite{He_2020_CVPR}, but adapt the learning rate scheduler (exponential scheduler) for better training convergence and keep shorter training (40 epochs) to limit over-fitting issues.

A first evaluation outcome, that comes from the direct observation of Fig.~\ref{fig_basic_models}, is that the assessment of the efficiency of the BFA only on the total number of bit-flips needed to reach a random-guess level is not an appropriate criterion. Indeed, after a fast decrease of the accuracy, the remaining efforts to reach the random-guess objective gather most of the bit-flips with the highest variance.  

We highlight the variability of the attack against the training initialization by reporting in~Table~\ref{BFA_adversary_goal} the performance of five models. For each model, we present the average number and the standard deviation of the bit-flips necessary to decrease the accuracy below 11\%, 25\%, 50\% and 75\%. The average is computed over five attacks. 
When the adversary's goal is set to 25\%, 50\% and 75\%, the standard deviation is limited, contrary to the random-guess objective (11\%). This important variability of the attack makes its evaluation more complex.

From an adversary point of view, we can question the necessity to performed so many faults to simply go from 25\% to 11\% while the model is no longer \textit{acceptable}. The random-guess objective is a radical threat model in which an adversary targets the integrity as well as the availability  of the model (i.e., the model is \textit{useless}) without any assumption on his real capacity. Indeed, typical threat models (for example, as defined in the majority of adversarial examples works) also consider an \textit{adversary's budget} that we claim to be an essential factor when dealing with fault injection attacks from RowHammer to laser beam injection.

\noindent\textbf{Takeaway \& Recommendation:} Practically, most of the fault injection attacks rely on a limited number of faults (for practical issues related to the injection mean as well as the presence of detection-based protections). We consider that fixing an adversary budget or, at least, using several gradual objectives is compulsory to properly evaluate the BFA. 

\section{Impact of training parameters}
\label{training_param}
Because of the very nature of BFA (alter the internal parameters of a model), the evaluation and analysis of the natural robustness of models with respect to their training is an essential step.
If some training parameters influence the BFA results, it becomes compulsory to take these factors into account when evaluating a defense, especially if the benefit offered by the defense is at the same level as the model's variability when trained with different parameters. Moreover, if some training parameters can make models more robust against BFA, it may help to promote \textit{good practice} to design and develop more secure models and, additionally, it may give some evidences for defense schemes.
Here, we investigate two training parameters that impact the choice of their initial distribution and the learning rate. 

The accuracy of the models used in these experiments are detailed in Table~\ref{tab_accuracy_cifar_0-1}, \ref{tab_accuracy_mnist_0-01}, \ref{tab_accuracy_cifar_0-01} (Appendix).

\subsection{Setups}
Experiments are conducted with VGG-11, ResNet-20 on CIFAR-10 and MLP on MNIST with two learning rates, $\lambda=0.1$ and $0.01$. Except learning rate and epochs\footnote{$\lambda=0.01$: 25 epochs for VGG-11, 40 epochs for ResNet-20 and MLP. $\lambda=0.1$:  40 epochs for VGG-11, MLP and ResNet-20.}, all training parameters are the same as in Section~\ref{discussion_threat_model}. 

\subsection{Learning rate}
\label{learning_rate}
We evaluate the impact of the learning rate $\lambda$ by using two initial values: 0.1 and 0.01. The scheduler is the same as before  (exponential scheduler - 0.95 - and a weight decay of $3.10^{-4}$). The weights are initialized using a normal distribution.

\begin{figure*}[t]
    \centering
    \setkeys{Gin}{width=0.49\textwidth} % <---
\subfloat[VGG-11, ResNet-20 on CIFAR-10 
\label{fig:learning_rate_cifar10}]{\includegraphics{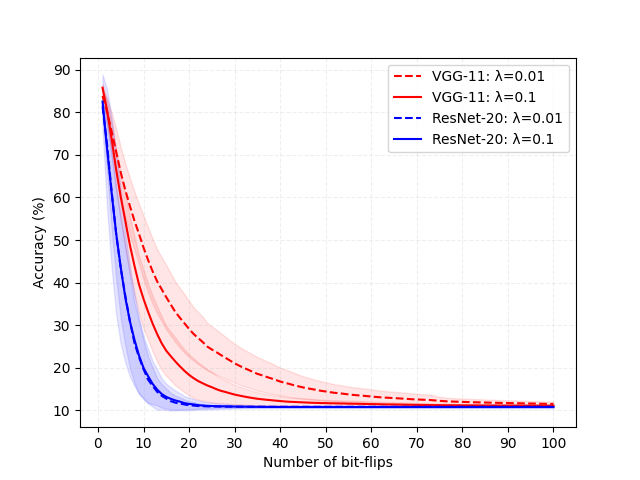} }\hfil
\subfloat[MLP on MNIST 
\label{fig:learning_rate_mnist}]{\includegraphics{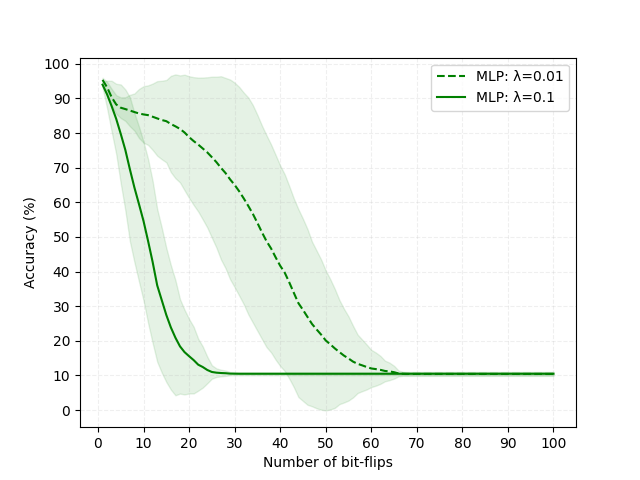} }\hfil
\vspace{15pt}
\subfloat[Gradient distribution with two $\lambda$, MLP (MNIST).    \label{fig:lambda_MLP}]{\includegraphics[scale=0.95]{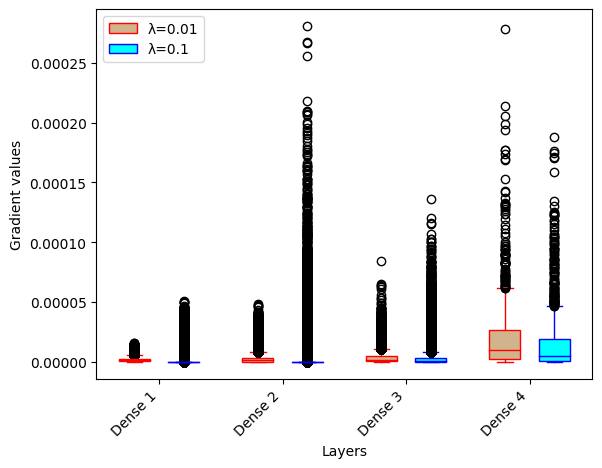} }
\caption{BFA performance with different learning rates (VGG-11 and ResNet-20 on CIFAR-10 and MLP on MNIST).}
\label{fig:learning_rate}
\end{figure*}

Fig.~\ref{fig:learning_rate} shows very opposite impact of $\lambda$ according to the architecture. We observe no influence of $\lambda$ on ResNet-20 whatever the adversary's objective. For VGG-11, the lowest $\lambda$ provides more robustness when the objective is to drop the accuracy below 40\% (a difference of almost 20 bit-flips is needed for 20\%). The most important influence is measured for MLP with a very significant difference of the number of bit-flips to reach the random-guess level (about 45 bit-flips).

We led further experiments for MLP by analyzing the distribution of the bit-flips and the gradients across the layers. Table~\ref{tab_distrib_mlp} shows how much the learning rate affects the bit-flips distribution. For $\lambda=0.01$, all the bit-flips are focused on the last layer. The distribution is more balanced for $\lambda=0.1$. These observations are confirmed by the gradients ($|\nabla_{W}\mathcal{L}|$) distribution for $\lambda=0.1$, with the highest gradients spread on the second, third and last layers (Fig.
\ref{fig:lambda_MLP}).

\begin{table}[htbp]
\caption{Learning rate ($\lambda$) influence: Bit-flips distribution and contribution per layer (MLP) for a random-guess goal.}
\begin{center}
\begin{tabular}{ccccc}
\toprule
\multirow{2}*{Layer} 
 & \multicolumn{2}{c}{$\lambda=0.01$} & \multicolumn{2}{c}{$\lambda=0.1$} \\
& bit-flips (\%) & Damage (\%) & bit-flips (\%) & Damage (\%) \\
%& bit-flips (\%) & & bit-flips (\%) & \\
\midrule
Dense 1 & 0 & 0 & 0 & 0 \\
Dense 2 & 0 & 0 & \textbf{56} & \textbf{66.4} \\
Dense 3 & 0 & 0 & 20 & 12.53 \\
Dense 4 & \textbf{100} & \textbf{100} & 24 & 21.07 \\
\bottomrule
\end{tabular}
\label{tab_distrib_mlp}
\end{center}
\end{table}

\subsection{Weights initialization}
\label{weights_initialization}
\begin{figure}[t!]
\centerline{\includegraphics[width=0.55\textwidth]{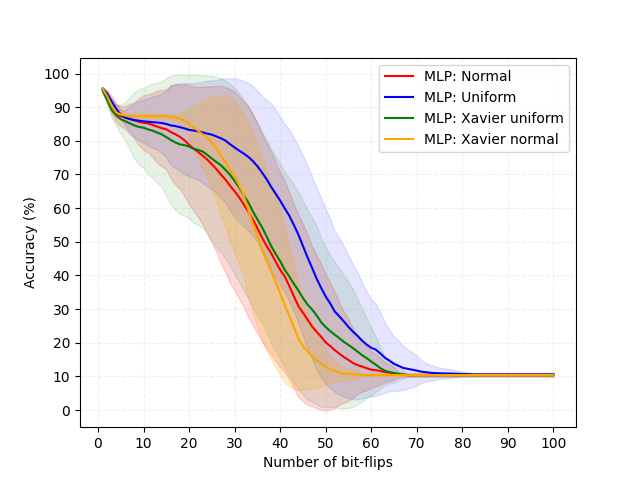}}
\caption{Impact of weight initialization for MLP (MNIST).}
\label{fig_init_weights_mlp}
\end{figure}

We fix the learning rate to $\lambda=0.01$ and use a Normal and a Uniform distribution for the weights initialization and their Xavier variants (that enables to keep the variance the same across every layers) ~\cite{glorot2010understanding}. Fig.~\ref{fig_init_weights_vgg11-resnet20} in the Appendix illustrates the impact of the initialization for both CNN architectures. For VGG-11 and ResNet-20, we do not observe a significant difference in the BFA behavior between those initializations. At most, a shift of 10 bit-flips is noted for MLP (see Fig.~\ref{fig_init_weights_mlp}) to the advantage of a Uniform distribution. However, as for the learning rate, this parameter should be properly noticed in evaluation reports for better comparison and reproducibility purpose.

\textbf{Takeaway \& Recommendation:} Depending on the architecture of the model, the training parameters may have a strong influence on the impact of the BFA and, therefore, should be carefully set and reported when evaluating the model's robustness. Analysis of the weights and the gradients distribution are efficient tools to better understand the model's behavior and explain potential variability of the attack.

\section{A Focus on Multilayer Perceptrons}
\label{focus_mlp}
Previous sections demonstrate that fully-connected models do not have the same behavior when facing BFA than CNN architectures (predominantly used in literature). In this section, we go deeper in the analysis by comparing MLP and CNN models as well as introducing the C-CNN model that is MLP with one additional convolution layer.

\subsection{Gradient distribution}

We observe that both VGG-11 and ResNet-20 models concentrate highest gradients in the first layers, as reported in previous works. We remind that, for a convolution layer, there are less parameters than a classical fully-connected layer since weights (that are the values of the convolutional filters) are shared. As mentioned in~\cite{Rakin_2019_ICCV} or \cite{He_2020_CVPR}, the error induced by bit flips performed on first convolution layers are accumulated and propagated throughout network. This error accumulation and propagation, exactly as for adversarial examples, explains the efficiency of BFA on such deep networks. 

As observed in Fig~\ref{fig:lambda_MLP}, the gradients distribution is significantly different for MLP with most of highest gradients at the model's end. 

Another important consequence is related to the gradient back-propagation as described in the Equation~\ref{eq_backprop}: a change (consequently to a bit-flip) of the value of the parameter at a layer $l$ directly alters the value of the gradients of the previous layers. This back-propagation phenomenon does not occur (or is limited) for bit-flips targeting the first layers which is not the case for the MLP since the highest gradients are located at the end of the network. For the case of an inner neuron $j$, the relation between a weight value and gradients is:    

\begin{equation}
\label{eq_backprop}
    \frac{\partial{\mathcal{L}}}{\partial{\mathbf{w_{i,j}}}} = a_i\rho_j \text{, with: } \rho_j = \Big(\sum_{k \in (l+1)}{\mathbf{w_{j,k}}\rho_k}\Big)\phi'(o_j) 
\end{equation}

$w_{i,j}$ is the weight that connects the neuron $j$ of the layer $l$ and the neuron $i$ of the previous layer ($l-1$), $a$ is the activation mapping of the neuron's output $o$ by the activation function $\phi$: $a_j=\phi(o_j)$ with $o_j=\sum_{k}{w_{k,j}a_k}$. A bit-flip on $w_{j,k}$ will change the value of $\nabla_{w_{i,j}}\mathcal{L}$. 

\subsection{Attack limitation}

\begin{figure}[b]
\centerline{\includegraphics[width=0.49\textwidth]{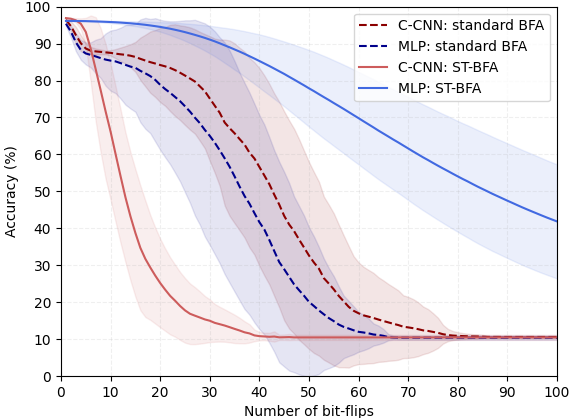}}
\caption{BFA results (C-CNN and MLP)}
\label{fig_BFA_results_cnnmlp}
\end{figure}

The previous observations raise an open question about the way BFA selects the most appropriate bits. PBS works well when all the highest gradients are concentrated on the earliest convolutional layers. In other cases, the PBS is not able to evaluate if a combination of bit-flips (that may benefit from error propagation) is more efficient than a single bit-flip yet associated to the highest gradient at the end of the model.

\begin{figure*}[h!]
    \centering
    \setkeys{Gin}{width=0.49\linewidth} % <---
\subfloat[C-CNN ($\lambda=0.01$)    \label{fig:CS_CCNN}]{\includegraphics{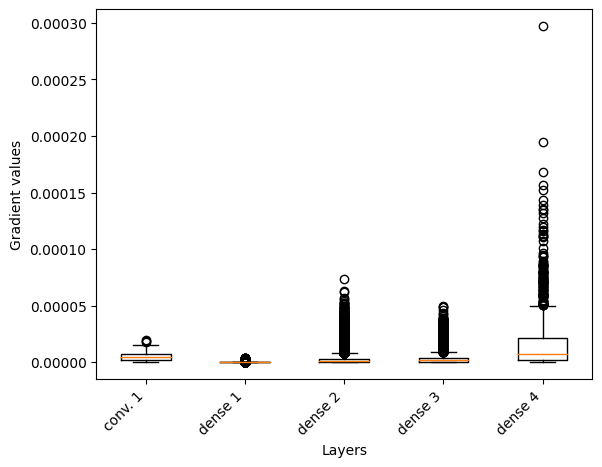}}
\hfil
\subfloat[C-CNN ($\lambda=0.01$), after 10 bit-flips.    \label{fig:CS_CCNN_after10}]{\includegraphics{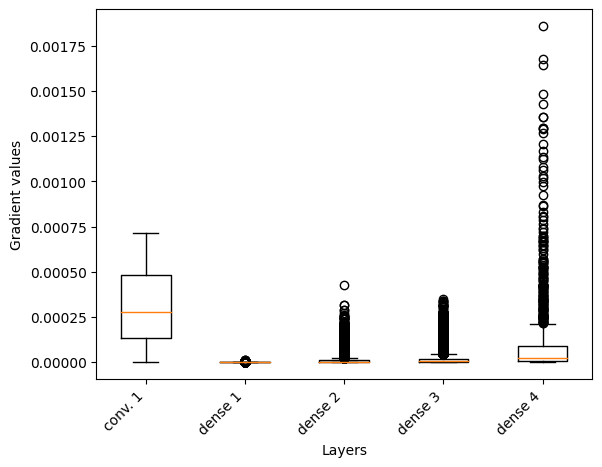} }
\caption{Gradient distribution for C-CNN before the $1^{st}$ (left) and after the $10^{th}$ (right) bit-flip. All the 10 bit-flips target the last fully-connected layer (\textit{dense 4}). After 10 bit-flip, gradients of the conv layer are significantly higher. }
\label{fig:gradient_distribution_ccnn}
\end{figure*}

To illustrate that potential limitation, we simply add a convolutional layer at the beginning of the MLP model. The resulting model, called C-CNN, has a gradient distribution that is very close to the original MLP (Fig.~\ref{fig:CS_CCNN} and~\ref{fig:lambda_MLP} respectively). Coherently, BFA has a relatively close performance against both models as illustrated in Fig.~\ref{fig_BFA_results_cnnmlp} (dotted lines) because the bit-flips (after PBS) will be almost exclusively concentrate on the last layer. 

Both models seem similar but what happen if we simply constraint the attack so that only the weights from the first layer are attacked.
This attack, noted ST-BFA (Spatially-Targeted BFA), provides a surprising result (red and blue lines in Fig.~\ref{fig_BFA_results_cnnmlp}). For C-CNN, the ST-BFA is far more efficient than the BFA: there is a difference of more than 50 bit-flips to reach a 20\% goal in favor of the ST-BFA. Figure~\ref{fig:CS_CCNN_after10} shows the gradients distribution after 10 bit-flips (that target exclusively the last layer) and demonstrates how much the bit-flips alter the gradients of the previous layers: the gradients of the convolutional first layer significantly increase during the attack (see the gradient level of \textit{conv1} in Fig.~\ref{fig:CS_CCNN} and Fig.~\ref{fig:CS_CCNN_after10}). 

Evaluating the C-CNN model with the standard BFA leads to a false robustness level since a stronger attack can be performed by only selecting one layer rather than the whole model.
On the contrary, for MLP, the ST-BFA has an opposite effect: the difference of bit-flips to reach 50\% of accuracy is closed to 40 bit-flips in favour of the standard BFA. 

\textbf{Takeaway \& Recommendation:} The results obtained on standard CNN architectures (ResNet, VGG) cannot be generalized to other architectures such as the multilayer perceptrons because of strong differences on the way gradients are distributed throughout the model. The BFA relies on a forward error accumulation and propagation as well as a backward propagation on the gradients. That results in situations where the BFA is significantly non-optimal compared to a localised application of the same attack. Therefore, to avoid evaluations that lead to a false sense of robustness, standard and localised-attacks should be carefully evaluated.

\section{Conclusion}
Parameter-based attacks are worrying threats that extent the attack surface of embedded neural networks models.
The Bit-Flip Attack is a state-of-the-art threat, that has been practically demonstrated with RowHammer attack (DRAM).
As for inference API-based attacks (adversarial examples) an important need is robust evaluation methodologies to properly assess the real impact of an attack as well as the level of robustness offered by defense schemes. We show that the standard threat model suffers from the lack of an adversary's bugdet while it is an important factor when dealing with fault injections. The random-guess goal is too radical since the BFA suffers from high variability when trying to reach this objective. We point out that this variability is also observed with training parameters as well as regarding the model's architecture. Therefore, the analysis of the weights and the gradients distribution appear as useful tools to better understand the mechanism of the BFA or to detect \textit{special cases} that could lead to a false sense of security. Finally, thanks to first experiments on pure fully-connected networks~--~that present very different behaviors than the models studied in the previous works~--~we show that the standard BFA could be significantly sub-optimal (focus the attack on only one layer leads to a 50 bit-flips gain on a model mixing convolutional and fully-connected layers), which highlights the need of careful, complete evaluations.

\section*{acknowledgement}
This work benefited from the French Jean Zay supercomputer with the AI dynamic access program.
This collaborative research is supported by (CEA-Leti) the European project ECSEL InSecTT\footnote{\url{www.insectt.eu}, InSecTT: ECSEL Joint Undertaking (JU) under grant agreement No 876038. The JU receives support from the European Union’s Horizon 2020 research and innovation program and Austria, Sweden, Spain, Italy, France, Portugal, Ireland, Finland, Slovenia, Poland, Netherlands, Turkey. The document reflects only the author’s view and the Commission is not responsible for any use that may be made of the information it contains.} and by the French National Research Agency (ANR) in the framework of the \textit{Investissements d’avenir} program (ANR-10-AIRT-05, irtnanoelec);  and (Mines Saint-Etienne) by the French program ANR PICTURE (AAPG2020).

\bibliographystyle{unsrt}
\bibliography{bib/biblio}

\clearpage
\newpage
\section*{Appendix}
\subsection{Models \& Training Parameters}
\label{annex_training_setup}
%%%%%%%%%%%%%%%%%%%%%%%%%%%%%%%%%%%%%%%%%%%%%%%%%%%%%%%%%%%%%%%%%%%%%%%%%%

Models and setups are as follow: 
\begin{table}[h!]
\caption{Section~\ref{standard_bfa_perf} Standard BFA performance. Same as \cite{He_2020_CVPR} }
\begin{center}
\begin{tabular}{ccc}
\toprule
Parameters&VGG-11&ResNet-20 \\
BatchNorm & Yes & Yes \\
Dropout & 0.2 & No \\
Weights Init & \multicolumn{2}{c}{Normal}\\
Optimizer & \multicolumn{2}{c}{SGD} \\
Learning rate & \multicolumn{2}{c}{$\lambda=0.1$} \\
Scheduler& \multicolumn{2}{c}{$\lambda/10$ at 80 and 120} \\
Epochs& \multicolumn{2}{c}{160} \\
Batch Size & \multicolumn{2}{c}{128} \\
\bottomrule
\end{tabular}
\label{tab_subsection3}
\end{center}
\end{table}

\begin{table}[h!]
\caption{Section~\ref{discussion_threat_model}. Adversary's goal \& budget}
\begin{center}
\begin{tabular}{ccc}
\toprule
Parameters&VGG-11&ResNet-20 \\
BatchNorm & No & Yes \\
Dropout & 0.5 & No \\
Weights Init & \multicolumn{2}{c}{Normal}\\
Optimzer & \multicolumn{2}{c}{SGD} \\
Learning rate & \multicolumn{2}{c}{$\lambda=0.1$} \\
Scheduler& \multicolumn{2}{c}{exponential (0.95)} \\
Epochs& \multicolumn{2}{c}{40} \\
Batch Size & \multicolumn{2}{c}{128} \\
\bottomrule

\end{tabular}
\label{tab_subsection4}
\end{center}
\end{table}

\begin{table}[h!]
\caption{Section~\ref{learning_rate} Learning Rate: $\lambda=0.1$, ResNet, VGG}
\begin{center}
\begin{tabular}{ccc}
\toprule
Parameters&VGG-11&ResNet-20 \\
BatchNorm & No & Yes \\
Dropout & 0.5 & No \\
Weights Init & \multicolumn{2}{c}{Normal}\\
Optimzer & \multicolumn{2}{c}{SGD} \\
Learning rate & \multicolumn{2}{c}{$\lambda=0.1$} \\
Scheduler& \multicolumn{2}{c}{exponential (0.95)} \\
Epochs& \multicolumn{2}{c}{40} \\
Batch Size & \multicolumn{2}{c}{128} \\
\bottomrule
\end{tabular}
\label{tab_subsection5_lambda_cnn_0_1}
\end{center}
\end{table}

\begin{table}[h!]
\caption{Section~\ref{learning_rate} Learning Rate: $\lambda=0.01$, ResNet, VGG}
\begin{center}
\begin{tabular}{ccc}
\toprule
Parameters&VGG-11&ResNet-20 \\
BatchNorm & No & Yes \\
Dropout & 0.5 & No \\
Weights Init & \multicolumn{2}{c}{Normal}\\
Optimzer & \multicolumn{2}{c}{SGD} \\
Learning rate & \multicolumn{2}{c}{$\lambda=0.01$} \\
Scheduler& \multicolumn{2}{c}{exponential (0.95)} \\
Epochs& 25 & 40 \\
Batch Size & \multicolumn{2}{c}{128} \\
\bottomrule
\end{tabular}
\label{tab_subsection5_lambda_cnn_0_01}
\end{center}
\end{table}

%%%%%%%%%%%%%% MLP %%%%%%%%%%%%%%
\begin{table}[h!]
\caption{Section~\ref{learning_rate}. Set-up for the learning rate experiments (MLP) $\lambda=0.1$ and $0.01$.}
\begin{center}
\begin{tabular}{cc}
\toprule
Parameters&MLP \\
BatchNorm & No  \\
Dropout & No \\
Weights Init & Normal\\
Optimzer & SGD \\
%Learning rate & $\lambda=0.1$ \\
Scheduler& exponential (0.95) \\
Epochs& 40 \\
Batch Size & 128 \\
\bottomrule
\end{tabular}
\label{tab_subsection5_lambda_mlp_0-1_0-01}
\end{center}
\end{table}

\begin{table}[h!]
\caption{Section~\ref{weights_initialization} Weight Initialization}
\begin{center}
\begin{tabular}{cccc}
\toprule
Parameters&VGG-11&ResNet-20&MLP \\
BatchNorm & No & Yes & No \\
Dropout & 0.5 & No & No\\
Weights Init & \multicolumn{2}{c}{Normal, Uniform, Xavier Normal, Xavier Uniform}\\
Optimzer & \multicolumn{2}{c}{SGD} \\
Learning rate & \multicolumn{2}{c}{$\lambda=0.01$} \\
Scheduler& \multicolumn{2}{c}{exponential (0.95)} \\
Epochs& 25 &40 &40 \\
Batch Size & \multicolumn{2}{c}{128} \\
\bottomrule
\end{tabular}
\label{tab_subsection5_weight_init}
\end{center}
\end{table}

\begin{table}[h!]
\caption{Section~\ref{focus_mlp} A Focus on MLP}
\begin{center}
\begin{tabular}{ccc}
\toprule
Parameters&MLP&C-CNN \\
BatchNorm & No & No \\
Dropout & No & No \\
Weights Init & \multicolumn{2}{c}{Normal}\\
Optimzer & \multicolumn{2}{c}{SGD} \\
Learning rate & \multicolumn{2}{c}{$\lambda=0.01$} \\
Scheduler& \multicolumn{2}{c}{exponential (0.95)} \\
Epochs& \multicolumn{2}{c}{40} \\
Batch Size & \multicolumn{2}{c}{128} \\
\bottomrule
\end{tabular}
\label{tab_subsection6}
\end{center}
\end{table}

\clearpage
\newpage

\subsection{Models accuracy}
\label{annex_model_accuracy}
\begin{table}[h!]
\caption{standard models accuracy}
\begin{center}
\begin{tabular}{cccccc}
\toprule
Model & 1 & 2 & 3 & 4 & 5 \\
\midrule
VGG-11 w/o BN & 91.35 & 91.14 & 90.95 & 91.21 & 91.35 \\
VGG-11 w BN & 92.35 & 92.46 & 91.14 & 92.34 & 92.13\\
ResNet-20 & 92 & 92.24 & 91.96 & 91.71 & 92.5 \\
\bottomrule
\end{tabular}
\label{tab_accuracy}
\end{center}
\end{table}

\begin{table}[h!]
\caption{Models accuracy with $\lambda = 0.1$ (CIFAR-10)}
\begin{center}
\begin{tabular}{ccccccc}
\toprule
Model & init Weights & 1 & 2 & 3 & 4 & 5 \\
\midrule
VGG-11 & Normal & 88.2 & 88.32 & 87.46 & 87.8 & 87.87 \\
ResNet-20 & Normal & 88.16 & 87.9 & 88.11 & 87.98 & 88.62\\
\bottomrule
\end{tabular}
\label{tab_accuracy_cifar_0-1}
\end{center}
\end{table}

\begin{table}[h!]
\caption{Models accuracy with $\lambda = 0.01$ (MNIST)}
\begin{center}
\begin{tabular}{ccccccc}
\toprule
Model & init Weights & 1 & 2 & 3 & 4 & 5 \\
\midrule
MLP & Normal & 96.17 & 95.96 & 96.25 & 96.07 & 96.16 \\
MLP & Uniform & 96.03 & 96.17 & 96.08 & 96.24 & 96.19\\
MLP & Xavier Normal & 96.15 & 96.3 & 96.22 & 96.26 & 96.11 \\
MLP & Xavier Uniform & 96.42 & 96.07 & 96.37 & 96.21 & 96.22 \\
C-CNN & Normal & 96.99 & 97.01 & 96.14 & 96.87 & 96.81 \\
\bottomrule
\end{tabular}
\label{tab_accuracy_mnist_0-01}
\end{center}
\end{table}

\begin{table}[h!]
\caption{Models accuracy with $\lambda = 0.01$ on (CIFAR-10)}
\begin{center}
\begin{tabular}{ccccccc}
\toprule
Model & init Weights & 1 & 2 & 3 & 4 & 5 \\
\midrule
VGG-11 & Normal & 85.08 & 85.19 & 85.7 & 85.6 & 85.47 \\
VGG-11 & Xavier Normal & 86.22 & 85.56 & 85.69 & 85.6 & 85.55 \\
VGG-11 & Xavier Uniform & 85.46 & 85.47 & 85.23 & 85.47 & 84.59\\
ResNet-20 & Normal & 84.84 & 85.95 & 85.2 & 85.04 & 84.81\\
ResNet-20 & Xavier Normal & 86.75 & 86.0 & 86.43 & 86.34 & 86.0 \\
ResNet-20 & Xavier Uniform & 86.56 & 86.76 & 85.91 & 86.93 & 86.14 \\
\bottomrule
\end{tabular}
\label{tab_accuracy_cifar_0-01}
\end{center}
\end{table}

\subsection{Bit-Flips Distribution}
%%%%%%%%%%%%%%%%%%%%%%%%%%%%%%%%%%%%%%%%%%%%%%%%%%%%%%%%%%%%%%%%%%%%%%%%%%
\begin{table}[h!]
\caption{VGG-11: Bit-Flip distribution and contribution}
\centering
\begin{tabular}{ccc}
\toprule
Layer & Dist. bit-flips (\%)&  Damage (\%) \\
\midrule
Conv. 1 & \textbf{22.27} & \textbf{50.71} \\
Conv. 2 & 15.38 & 25.64 \\
Conv. 3 & 14.47 & 10.75 \\
Conv. 4 & 13.89 & 5.11 \\
Conv. 5 & 12.94 & 4.9 \\
Conv. 6 & 6.24 & 0.89 \\
Conv. 7 & 2.06 & 0.01\\
Conv. 8 & 0 & 0 \\
Denses & 12.75 & 1.99 \\
\bottomrule
\end{tabular}
\label{tab_distrib_vgg11}
\end{table}

\begin{table}[h!]
\caption{ResNet-20: Bit-Flip distribution and contribution}
\centering
\begin{tabular}{ccc}
\toprule
Layer & Dist. bit-flips (\%)& Damage (\%) \\
\midrule
Conv. 1 & 21.35 & 23.87 \\
Stage 1.0 & \textbf{14.16} & \textbf{21.49} \\
Stage 1.1 & 11.76 & 27.0 \\
Stage 1.2 & 2.4 & 0.61 \\
Stage 2.0 & 8.5 & 7.12 \\
Stage 2.1 & 8.28 & 7.41 \\
Stage 2.2 & 7.18 & 6.01 \\
Stage 3.0 & 3.27 & 0.73 \\
Stage 3.1 & 3.49 & 0.2 \\
Stage 3.2 & 14.81 & 4.62 \\
Dense 1 & 4.80 & 0.95 \\
\bottomrule
\end{tabular}

\label{tab_distrib_resnet20}
\end{table}

%%%%%%%%%%%%%%%%%%%%%%%%%%%%%%%%%%%%%%%%%%%%%%%%%%%%%%%%%%%%%%%%%%%%%%%%%%

\newpage
\subsection{Weights distribution (\textbf{Section~\ref{weights_initialization}})}
\label{appendix_wd}
%%%%%%%%%%%%%%%%%%%%%%%%%%%%%%%%%%%%%%%%%%%%%%%%%%%%%%%%%%%%%%%%%%%%%%%%%%
Models follow the training setup from Table~\ref{tab_subsection3}. The quantization into 8-bit representation is performed as follows:

\begin{equation*}
    w^{q}=round\Big(\frac{hardtanh(w,-max(|W_l|),max(|W_l|)}{127}\Big)max(|W_l|))
\end{equation*}
With $w$ a weight belonging to $W_l$, the tensor of a layer $l$.

\begin{figure*}[h]
    \centering
    \setkeys{Gin}{width=0.49\textwidth} % <---

\subfloat[Model from \cite{He_2020_CVPR}\footnote{pretrained weights coming from \url{https://github.com/elliothe/BFA/issues/7}} 8-bit 
\label{fig:wd_71_8}]{\includegraphics{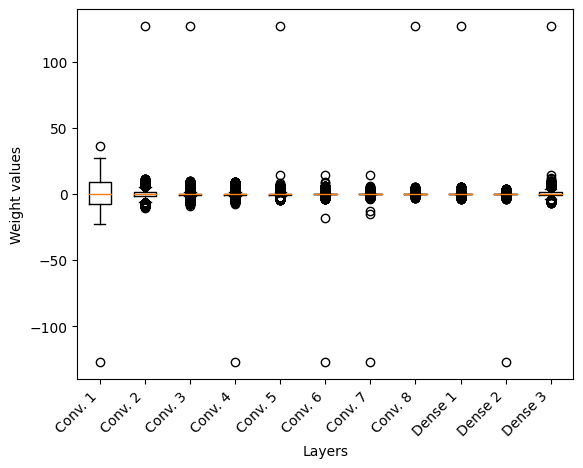} }\hfil
\subfloat[Model \#1 (ours) 8-bit 
\label{fig:wd_71_8}]{\includegraphics{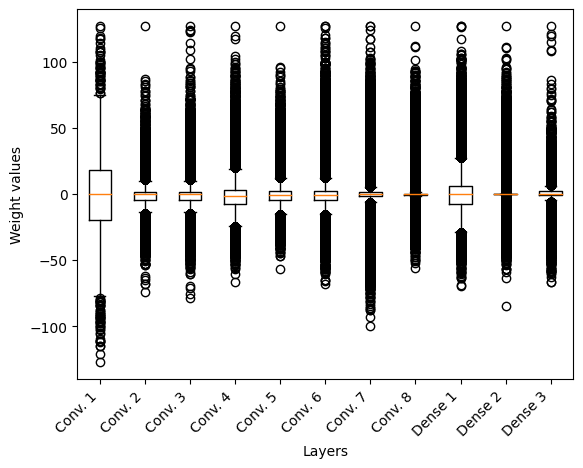} }\hfil
\subfloat[Model \#2 (ours) 8-bit 
\label{fig:wd_475_8}]{\includegraphics{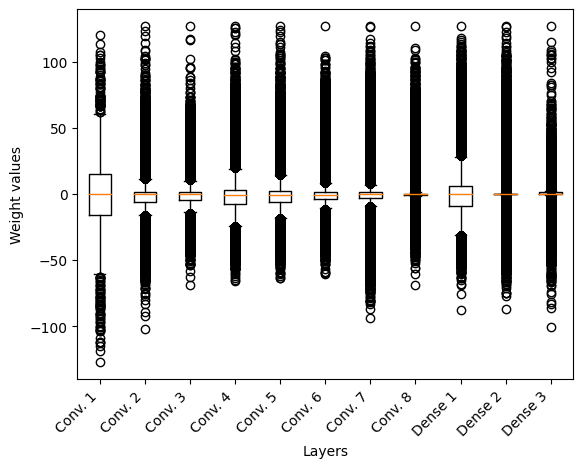} }\hfil
\subfloat[Model \#3 (ours) 8-bit 
\label{fig:wd_2476_8}]{\includegraphics{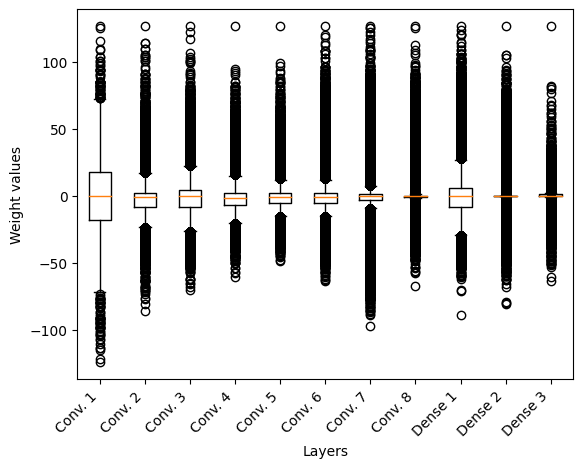} }\hfil
\subfloat[Model \#4 (ours) 8-bit
\label{fig:wd_3878_8}]{\includegraphics{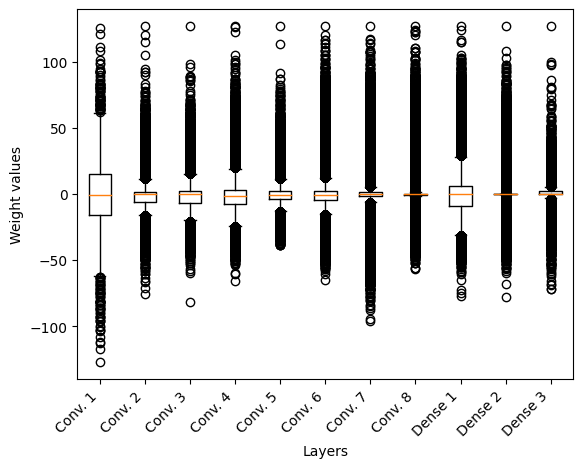} }\hfil
\subfloat[Model \#5 (ours) 8-bit
\label{fig:wd_8713_8}]{\includegraphics{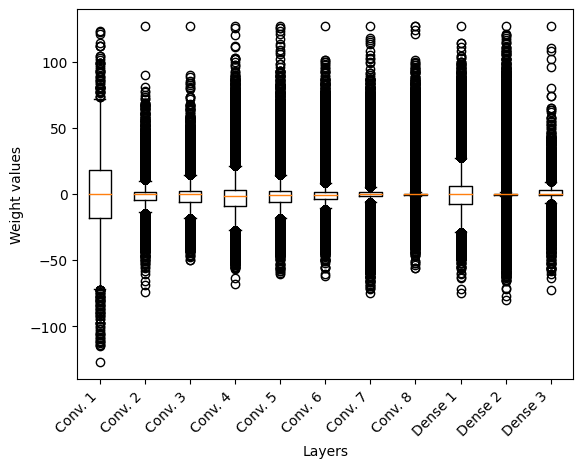} }\hfil
\caption{Weight distributions for VGG-11 models: from \cite{He_2020_CVPR} (top, left) and ours in 8-bit quantization.}
\label{fig_weigths_distribution_8bits}
\end{figure*}

The distribution of the weights of VGG-11 models are presented in Fig. \ref{fig_weigths_distribution_8bits}: 
\begin{itemize}
    \item VGG-11 proposed by the authors of~\cite{He_2020_CVPR} in \url{https://github.com/elliothe/BFA/issues/7} that is attacked with only 10 bit-flips. 
    \item Our VGG-11 model with 5 runs (5 training seeds).  
\end{itemize}

\subsection{Influence of the learning rate.}
\begin{figure*}[h!]
    \centering
    \setkeys{Gin}{width=0.49\textwidth} % <---
\subfloat[VGG11 ($\lambda=0.01$) \label{fig:GD_VGG_001}]{\includegraphics{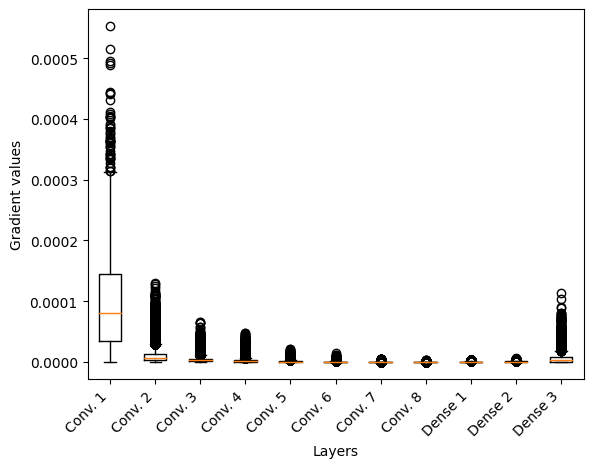} }\hfil
\subfloat[VGG11 ($\lambda=0.1$) \label{fig:GD_VGG_01}]{\includegraphics{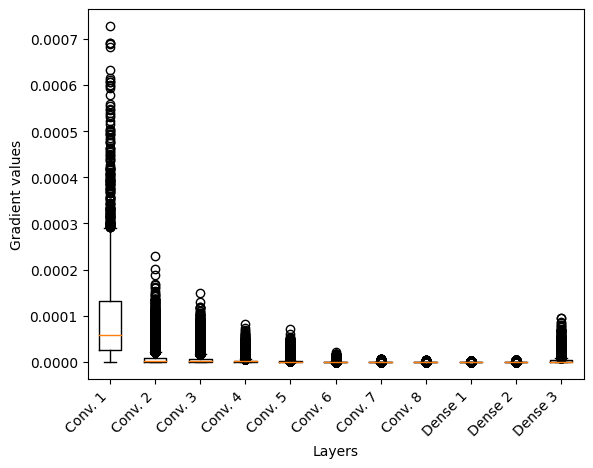} }\hfil
\caption{Gradient distributions for VGG-11}
\label{fig:GD_VGG}
%\vspace{-25pt}
\end{figure*}

\begin{figure*}[h]
    \centering
    \setkeys{Gin}{width=0.49\textwidth} % <---
\subfloat[ResNet-20 ($\lambda=0.01$) \label{fig:GD_RESNET_001}]{\includegraphics{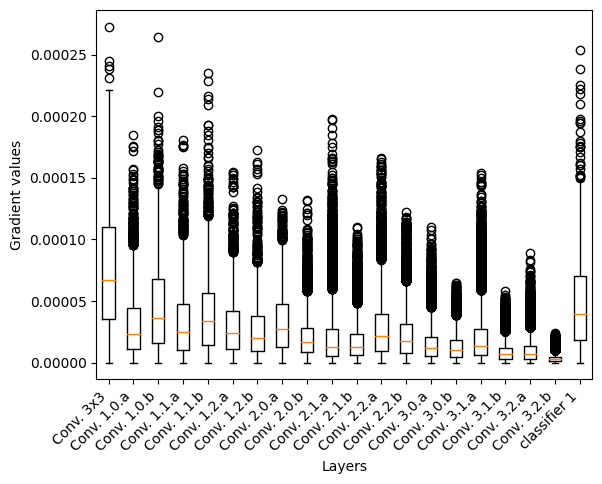} }\hfil
\subfloat[resNet-20 ($\lambda=0.1$) \label{fig:GD_RESNET_01}]{\includegraphics{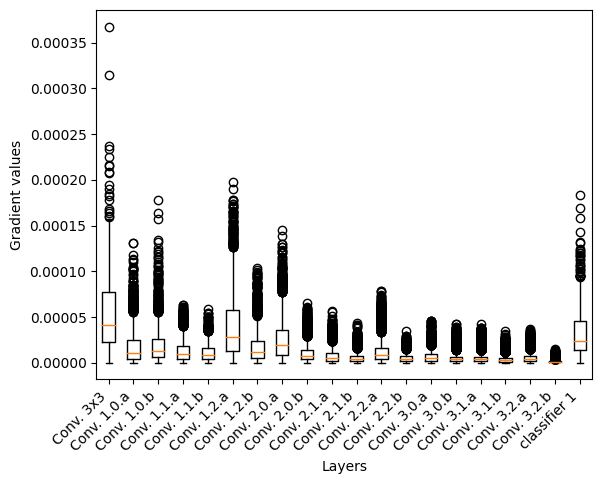} }\hfil
\caption{Gradient distributions for ResNet-20}
\label{fig:GD_RESNET}
\vspace{-15pt}
\end{figure*}

\begin{figure*}[h]
    \centering
    \setkeys{Gin}{width=0.49\textwidth} % <---
\subfloat[C-CNN ($\lambda=0.01$) \label{fig:GD_CCNN}]{\includegraphics{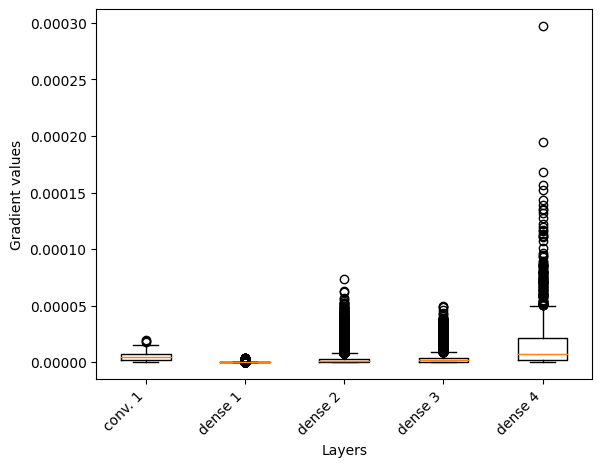} }\hfil
\subfloat[MLP ($\lambda=0.01$ and $\lambda=0.1$) \label{fig:GD_MLP}]{\includegraphics{imgs/gradient_distrib/grad_moustache_mlp_lr_all.png} }\hfil
\caption{Gradient distribution for C-CNN and MLP}
\label{fig:GD_CCNN_MLP}
\vspace{-20pt}
\end{figure*}

Fig.~\ref{fig:GD_VGG}, \ref{fig:GD_RESNET} and \ref{fig:GD_CCNN_MLP} show the distribution of the gradients with two learning rates $\lambda=0.1$ and $0.01$. Fig.~\ref{fig:learning_rate_cifar10} and Fig.~\ref{fig:learning_rate_mnist} show the impact of $\lambda$ on the BFA for VGG-11 and ResNet-20 on CIFAR10 and MLP on MNIST.  

The distribution of the gradients for the C-CNN illustrating the BFA limitation is presented in Fig.~\ref{fig:GD_CCNN}.

\subsection{Influence of weight initialization}
\begin{figure*}[t]
    \centering
    \setkeys{Gin}{width=0.49\textwidth} % <---
\subfloat[VGG-11 on CIFAR10 
\label{fig:init_weights_cifar10}]{\includegraphics{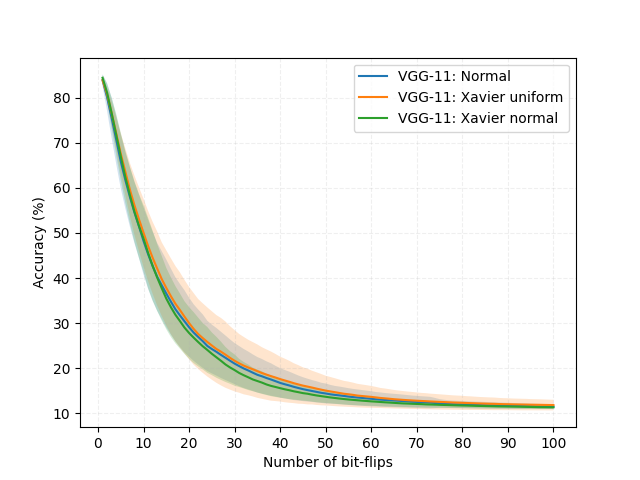} }\hfil
\subfloat[ResNet-20 on CIFAR10 
\label{fig:init_weights_mnist}]{\includegraphics{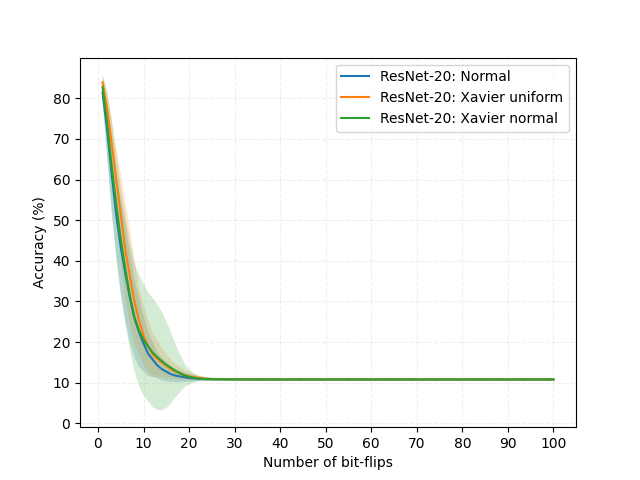} }\hfil
\caption{BFA performance with different weights initialization (VGG-11 and ResNet-20).}
\label{fig_init_weights_vgg11-resnet20}
\vspace{-20pt}
\end{figure*}

\begin{figure*}[t]
\centerline{\includegraphics[width=0.53\textwidth]{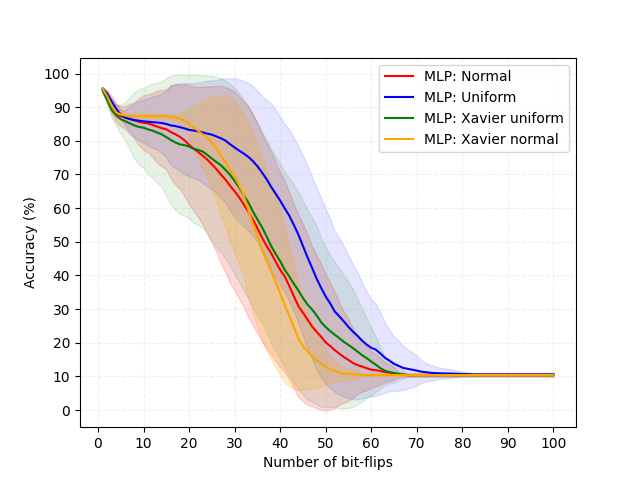}}
\caption{BFA performance with different weights initialization (MLP).}
\label{fig_init_weights_mlp1}
\vspace{-20pt}
\end{figure*}

Impact of the initialization of the weights is presented in Fig.~\ref{fig_init_weights_vgg11-resnet20} for VGG-11 and ResNet-20 on CIFAR10 and Fig.~\ref{fig_init_weights_mlp1} for MLP on MNIST. 

\clearpage
\newpage

\end{document}